# High gain metalens antenna transceiver for terahertz communication

Zebin Huang, Qun Zhang, Feifan Han, Hao Wang, Shuyi Chen, Weichao Li, Xiongbin Yu and Xiaofeng Tao, *Senior Member, IEEE*

*Abstract*—Terahertz (THz) metalens antennas with compact planar structures have demonstrated significant potential in enhancing gain and aperture efficiency through beam convergence. However, research on THz wireless communication systems utilizing metalens antennas remains limited, primarily due to insufficient collaborative enhancement in gain and bandwidth in THz transceiver design. In this paper, we propose a high gain metalens antenna transceiver and demonstrate its application for THz communication. The system employs a horn antenna integrated with a 3D-printed bracket to enhance the metalens gain and operating bandwidth, where the metalens adopts a "sandwich" architecture composed of a V-shaped copper resonator, a dielectric substrate, and a grating. The resonant design inside the metalens facilitates high polarization conversion efficiency and full phase modulation across a 0° to 360 ° range at frequency between 0.20 to 0.30 THz band. Experimental results demonstrate a peak gain of 36.1 dBi and aperture efficiency of 54.45% at 0.244 THz, with a 3 dB bandwidth exceeding 33 GHz. A prototype communication system incorporating the metalens transceiver achieves a bit error rate (BER) reduction by three orders of magnitude compared to conventional horn antennas and supports a maximum data rate of 100 Gbps. This proposed metalens offer a high-gain, compact solution for achieving high data rate THz communications, driving advancements in 6G communication network.

## I. Introduction

TERAHERTZ (THz) waves, typically spanning frequencies from 0.1 THz to 10 THz, exhibit immense potential for 6G communication networks due to their ultra-wideband capabilities, facilitating high data rates and low-latency interconnections [1], [2], [3], [4]. Despite these advantages, the practical implementation of THz waves in wireless communication is constrained by significant path loss and absorption effects, especially in the extreme transmission environment [5], [6], [7]. Therefore, the development of high gain, compact, and broadband antenna transceiver is essential to increase data rates in THz communication. Conventional polymer-based lenses rely on the geometrical design of curved surfaces or fabricating unit cells with different height to introduce phase delays, which further achieves beam convergence when THz wave transmitting through them [8]. However, this approach typically requires multiple lens sets or complicated fabrication procedure, resulting in a complex and bulky communication system that hinders the development of 6G THz communication networks.

Metasurfaces are two-dimensional artificial structures composed of subwavelength periodic meta-atoms (unit cells), offering precise control over electromagnetic waves through tailored geometry and material properties [9], [10], [11], [12], [13], [14], which have shown prospective applications in compact THz communication networks. These sub-wavelength artificial structures enable efficient and flexible control over incident waves' phase, amplitude, and polarization states by exploiting their unique responses [15]. The abilities of metasurfaces to facilitate electromagnetic-matter interactions have garnered significant attention, leading to the development of various THz meta-components, such as beam deflectors [16], [17], holograms [18], [19], [20], vortex generators [21], [22], [23], and metalenses [24], [25], [26], [27], [28].

THz metalens stands out as a pivotal application for 6G communication networks as it can increase communication performance by boosting the received gain of the antenna [29], [30], [31], [32]. Typically, the Fresnel zone plate, characterized by its binary concentric rings, provides a straightforward approach to manipulate THz waves through precise beam shaping and focusing by tailoring its ring pattern [24], [26], [33]. However, the binary design of these plates can introduce paraxial diffraction, potentially reduce the gain of metalens and limits to below 20 dBi and operation bandwidth below 10 GHz at 0.27 THz [24]. By employing the multi-layer metal resonance architecture, the 0 to 360° phase control can be achieved at the cross-polarization component and the gain of the metalens antennas can achieve 28.8 dBi with operation bandwidth of 28 GHz at 0.25 THz [27]. Moreover, the all-dielectric architecture can realize a better broadband effect and the phase delay is achieved through varying the height and materials of meta-atoms. The gain and bandwidth of all-

This work was supported in part by the Major Key Project of PCL (PCL2023AS1-3, 2024ZY1A0010) and in part by the China Postdoctoral Science Foundation Funded Project (2024M761543). (Zebin Huang and Qun Zhang contributed equally to this paper) (Corresponding author: Xiongbin Yu)

Zebin Huang, Qun Zhang, Feifan Han, Hao Wang, Shuyi Chen, Weichao Li, Xiongbin Yu are with the Department of Broadband Communication, Pengcheng Laboratory, Shenzhen, China. (email: {huangzb, zhangq07, hanff, wangh10, chenshy, yuxb, liwch}@pcl.ac.cn)

Xiaofeng Tao is with the National Engineering Research Center of Mobile Network Technologies, Beijing University of Posts and Telecommunications, Beijing, China and also with the Department of Broadband Communication, Pengcheng Laboratory, Shenzhen, China (email: taoxf@bupt.edu.cn).

Supplementary Video I is available online at: http://ieeexplore.ieee.org.

Color versions of one or more of the figures in this article are available online at http://ieeexplore.ieee.org

dielectric metalens reach 30.8 dBi and 55 GHz, respectively [8]. The fabrication challenges of all-dielectric metalens with varying height structures hinder the widespread adoption in compact THz wireless communication systems. Therefore, toward practical applications necessitate metalens antennas with collaborative enhancements in both gain and operations bandwidth, while achieving compact and economic solutions for fabrication.

In this paper, we propose and design a high gain metalens antennas transceiver and further demonstrate its applications for THz communication. The horn antenna with 3D-printed bracket mounted was employed for enhancing the gain of metalens. Metalens is constructed using a 'sandwich' structure comprising a V-shaped copper structure, dielectric layer, and grating, enabling high polarization conversion efficiency and full phase modulation across the 0° to 360° range via establishing a THz resonant structure among layers. The results demonstrate that the transmissive metalens with a focal distance of 25 mm achieves a maximum gain of 36.10 dBi and aperture efficiency of 54.45 % at 0.244 THz, with a 3 dB bandwidth of 33.0 GHz. Consequently, we proposed a compacted THz transceivers prototype system with metalens antennas transceiver, realizing reduction of bit error rate (BER) with three orders of magnitude and successfully transmitting 100 Gbps 32-quadrature amplitude modulation (QAM) signals with BER below $10^{-2}$. We demonstrate the application of real-time uncompressed 4K high-definition video transmission. These remarkable findings underscore the potential of metalens for establishing THz communication network, enabling high data rate and low-latency transmission, thereby advancing 6G wireless communications technologies.

The key contributions of our paper are as follows:

1) High gain THz transceiver: The transmissive metalens enables high conversion efficiency benefiting from the proper design of two-layer meta-atoms, reducing the insertion loss of THz transceiver. Concurrently, the capability for full phase manipulation effectively mitigates the paraxial diffraction of electromagnetic waves, thereby enhancing the realized gain of the metalens. Therefore, we propose the THz metalens transceiver with combination of metalens and standard horn antenna, achieves a peak gain of 36.1 dBi and aperture efficiency of 54.45 %, operating within a 3dB gain bandwidth exceeding 33.0 GHz.

2) Prototype system for high data rate THz communication: Exiting research concerning metalens in THz or microwave frequencies is primarily confined to the characteristics analysis of elements, such as insertion loss, gain, radiation patterns, and operating bandwidth. However, previous research has not demonstrated the THz communication link due to the lack of collaborative enhancement in gain and operating bandwidth. We carried out comprehensive analysis of metalens antennas together with experimental demonstrations in THz communication using metalens antenna transceiver, achieving the reduction of BER by three orders of magnitude and a maximum data rate of 100 Gbps. Consequently, we demonstrated the prototype system by transmitting real-time uncompressed 4K high-definition video, showcasing the feasibility of metalens antenna transceiver in THz communication.

The rest of this paper is structured as follows: Section II presents the methodologies and foundational principles of the metalens antennas transceiver, encompassing the characterization of meta-atoms using Jones matrices and the numerical analysis of these meta-atoms. Section III details the experimental procedures and outcomes, which include the characterization of the metalens using a vector network analyzer, the results from the THz communication link experiments using metalens antenna transceiver, and the demonstration of a prototype system for real-time 4K video transmission. Section IV evaluates the ultimate performance capabilities of the metalens and gives out the comparison of metalens in THz frequency ranges. Lastly, Section V provides the conclusions of this paper.

## II. METHODS AND PRINCIPLES

Figure 1 illustrates the schematic diagram of THz transceiver based on metalens antennas. Figure 1 (a) shows the overview of THz transceiver, including multiplier, sub-harmonic mixer, horn antennas and metalens. The local oscillator (LO) is multiplied by multiplier and then coupled with the immediate frequency (IF) signal in the sub-harmonic mixer, thus generating the THz electromagnetic waves with digital signals encoded. The metalens antenna is shown in Fig. 1(b), where the horn antenna is mounted on a 3D-printed bracket and the metalens is placed on the outer side of the bracket. Note that we employed two different types of horn antennas with gain of 15 and 23 dBi to generate THz electromagnetic waves with different waist radius. The architecture of the periodical arrangement of single meta-atom is detailed in Fig. 1(c). We propose a double-layered 'copper-dielectric-copper' architecture. The top layer features a V-shaped copper structure with varying arm lengths $L$ and opening angles $\theta$, while the bottom layer consists of a grating with a fixed period (100 $\mu$m in this case). This resonant structure can effectively improve the conversion efficiency of meta-atoms. The manipulation principle of the meta-atom is based on the interaction between the V-shaped structure and the periodic grating. The V-shaped structure functions as a half-wave plate, converting linearly polarized THz waves into cross-polarized states. Meanwhile, the periodic grating in the bottom layer acts as a linear polarizer, filtering out co-polarized states. As shown in Fig. 2(a), the metalens will focus THz electromagnetic waves on the focal plane with high gain achieved, while the polarization states will be converted after the metalens.

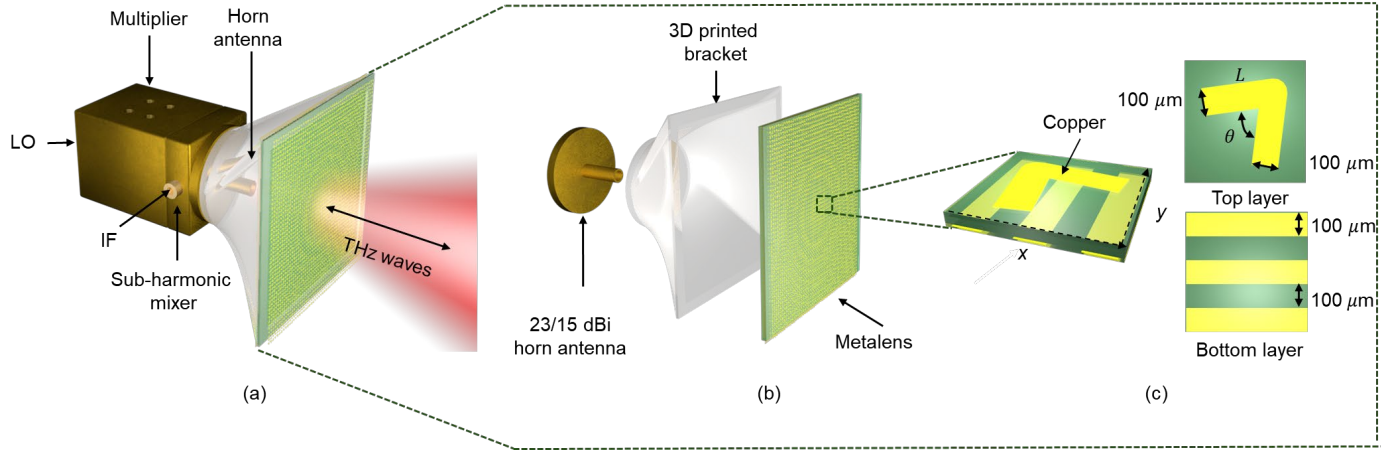

Fig.1 Schematic diagram of high gain THz transceiver using metalens antenna. (a) Overview of THz transceiver. LO: local oscillator; IF: intermediate frequency. (b) Diagram of metalens antennas mounted by a 3D-printed bracket. (c) Single meta-atom architecture of metalens.

Consequently, the Jones matrix for a meta-atom composed of the V-shaped structure and grating can be expressed as:

$$J_f = J_g \cdot J_v$$
$$= \begin{bmatrix} 1 & 0 \\ 0 & 0 \end{bmatrix} \begin{bmatrix} 0 & T_{xy}\exp(i\varphi_{xy}) \\ T_{yx}\exp(i\varphi_{yx}) & 0 \end{bmatrix} \quad (1)$$
$$= \begin{bmatrix} 0 & T_{xy}\exp(i\varphi_{xy}) \\ 0 & 0 \end{bmatrix},$$

where $T_{xy}$ represents the transmission ratio with y polarization state input and x polarization state output, $\varphi_{xy}$ represents the corresponding phase modulation. $J_g$ and $J_v$ represent the Jones matrices of grating and V-shaped structure, $J_f$ is the forward transmission Jones matrix. According to the reversibility of beam propagation, $T_{xy} = T_{yx}$ and $\varphi_{xy} = \varphi_{yx}$. For instance, a y-polarized THz wave incident on the meta-atom can be mathematically expressed as:

$$E_o = J_f \cdot \begin{bmatrix} 0 \\ 1 \end{bmatrix}$$
$$= \begin{bmatrix} 0 & T_{xy}\exp(i\varphi_{xy}) \\ 0 & 0 \end{bmatrix} \cdot \begin{bmatrix} 0 \\ 1 \end{bmatrix} \quad (2)$$
$$= \begin{bmatrix} T_{xy}\exp(i\varphi_{xy}) \\ 0 \end{bmatrix},$$

where the y-polarization state is converted to the x-polarization state with a corresponding phase modulation, and the periodical grating reflects the x-polarization state. Conversely, the backward Jones matrix can be expressed as:

$$J_b = J_v \cdot J_g$$
$$= \begin{bmatrix} 0 & T_{xy}\exp(i\varphi_{xy}) \\ T_{yx}\exp(i\varphi_{yx}) & 0 \end{bmatrix} \begin{bmatrix} 1 & 0 \\ 0 & 0 \end{bmatrix} \quad (3)$$
$$= \begin{bmatrix} 0 & 0 \\ T_{yx}\exp(i\varphi_{yx}) & 0 \end{bmatrix},$$

where $J_b$ is the backward transmission Jones matrix. The x-polarization incident from the grating side will be converted to y-polarization with corresponding phase modulations.

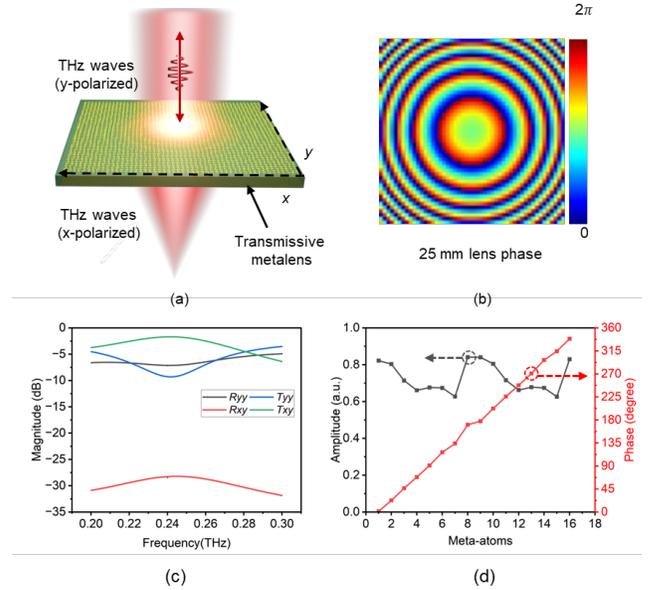

Fig.2 (a) Schematic diagram of THz wave transmission using metalens. (b-d) Numerical results of transmission and reflection coefficients for different unit structure of metalens at frequency ranging from 0.20 THz to 0.30 THz. (a) Transmission and reflection coefficients of unit structure with $L$= 500 um and $\theta = 75°$. (b) Phase and amplitude coefficient of $T_{xy}$ for all 16 units that covering 0 to $2\pi$ phase modulations at 0.24 THz.

Furthermore, the metalens is constructed by arranging a two-dimensional meta-atom distribution according to the desired phase modulation function. The phase distribution of the high-gain transmissive lens can be expressed as:

$$\varphi_c(x,y,f) = \frac{kx^2 + ky^2}{2f}, \quad (4)$$

where $(x, y)$ is the Cartesian coordinate, and $\varphi_c$ represents the convex phase distribution. $k = 2\pi / \lambda$ is the wave vector, $\lambda$ is the operating wavelength, and $f$ is the focal length of lens phase. The corresponding phase distribution with focal length of 25 mm is depicted on Fig. 2(b).

The electromagnetic properties of the meta-atoms were calculated using the finite-difference frequency-domain (FDTD) method. The V-shaped structure and grating structures are fabricated from copper with a thickness of 0.5 oz (~17 $\mu m$). To enhance the transmittance of the metalens, Rogers RT5880 was selected as the dielectric substrate, featuring a dielectric constant of 2.20 and a thickness of 127 $\mu m$. The periodicity of meta-atoms is set to 600 $\mu m$, approximately half the operating wavelength. The arm length $L$ of the meta-atoms is varied between 350 $\mu m$ and 550 $\mu m$, while the opening angles are configured to 30°, 45°, 60°, and 75°. The numerical calculation results are shown in Fig. 2(c) and (d). Figure 2(c) illustrates the simulated electromagnetic results of a selected meta-atom at frequency range between 0.20 to 0.30 THz, with an arm length of 500 $\mu m$ and an opening angle of 75°. For simplicity expression, $T_{xy}$ and $R_{yy}$ represent the transmission and reflection coefficients. The first sub-letter represents the received polarization state, and the second represents the transmitted polarization state. In this part, we determine the forward transmission with the y-polarization incident. According to the Jones matrix representation in Eq. (1–3), the incident y-polarized THz wave undergoes conversion to x-polarization with additional phase modulation. Notably, the polarization suppression ratio of these meta-atoms can achieve approximately 7.59 dB at 0.24 THz, with a 3 dB bandwidth spanning 70 GHz. Figure 2(d) shows the $T_{xy}$ phase and amplitude modulation of 16 selected meta-atoms. The results indicate that the selected structures effectively achieve approximate full phase modulation across the 0 to 360° segment, with transmittance exceeding 62%. In this work, the 0° and 360° share the same meta-atoms for phase modulations.

To enhance the reproducibility of this work, we provide detailed geometrical parameters and calculated electromagnetic responses of V-shaped structure, shown in Table I. The geometric configurations of all meta-atoms utilized in the design procedure are shown in Fig. 3(a), and the numerically calculated phase versus frequency response for all meta-atoms (calculated by frequency domain finite difference) are shown in Fig. 3(b), revealing a smooth phase progression as frequency increases between 0.20 and 0.30 THz. This characteristic indicates that these meta-atoms possess a relatively broad bandwidth response. Note that the calculated phase response may deviate from the target phase modulations, however, this may not significantly affect the results of metalens. The meta-atoms with phase modulation between 177.59° and 338.75° are derived by counterclockwise rotating the initial eight meta-atoms by 90°.

TABLE I
PARAMETERS OF META-ATOMS FOR FULL PHASE MODULATIONS AT 0.24 THZ.

| Phase (°) | Amp. (a.u.) | $L$ (um) | $\theta$ (°) | Phase (°) | Amp. (a.u.) | $L$ (um) | $\theta$ (°) |
|---|---|---|---|---|---|---|---|
| 1.41 | 0.82 | 530 | 75 | 177.59 | 0.84 | 520 | 45 |
| 22.65 | 0.80 | 500 | 75 | 202.32 | 0.80 | 500 | 75 |
| 46.4 | 0.71 | 460 | 75 | 225.78 | 0.71 | 460 | 75 |
| 68.29 | 0.66 | 420 | 75 | 247.78 | 0.66 | 420 | 75 |
| 90.95 | 0.67 | 390 | 75 | 270.74 | 0.67 | 390 | 75 |
| 116.70 | 0.67 | 370 | 75 | 297.04 | 0.67 | 370 | 75 |
| 134.04 | 0.62 | 360 | 75 | 314.24 | 0.62 | 360 | 75 |
| 159.54 | 0.82 | 530 | 45 | 338.75 | 0.82 | 550 | 45 |

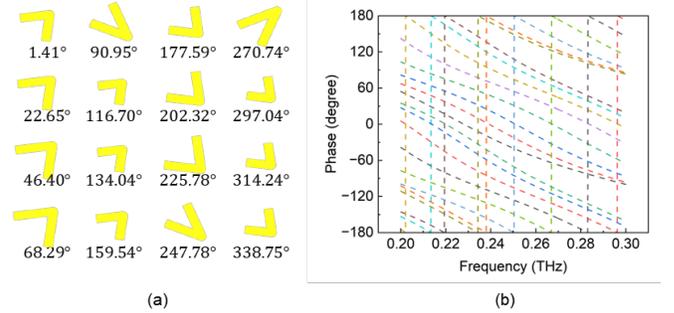

Fig. 3. Numerical results of meta-atom of metalens (a) Geometrical results of meta-atoms covering the modulation from 0 to 360°, with a design phase interval of approximately 22.5°. The 177.59° to 338.75° modulations were achieved by counterclockwise rotating the V-shaped structure with 90°. (b) The modulated phase values versus frequency from 0.20 to 0.30 THz for all meta unit calculated by FDTD.

## III. RESULTS AND ANALYSIS

### A. Measurement of THz metalens using vector network analyzer.

The experiment setup of the THz metalens measurement system is depicted in Fig. 4(a). This involves a vector network analyzer, metalens antennas (combined with horn antenna), and 4- axis electronic displacement platforms. The vector network analyzer is utilized to analyze S-parameters of the fabricated metalens, which is equipped with two frequency expanders, which enables the analysis of terahertz waves with frequencies ranging from 0.17 THz to 0.26 THz. In this paper, we utilized two standard conical horn antennas with aperture diameter of 7.8 mm (23 dBi gain) and 3.6 mm (15 dBi gain), thereby generating electromagnetic waves with different waist radius for comparing the experimental results of metalens for communication.

To increase the experiment feasibility, the size of metalens is set to be 30 mm × 30 mm (50 × 50 units), and the photograph of metalens mounted on a 3D-printed bracket is shown in Fig. 4(c). Consequently, the standard circular horn antennas are mounted on the WR-4 waveguide ports of the transmitter and receiver. The transmitter and receiver are affixed to 4-axis electronic displacement platforms, which facilitate the scanning of radiation patterns. Since the metalens will convert the polarization states during phase modulation, a

90° twisted waveguide (1-inch, 25.4 mm) should be mounted before waveguide ports for polarization matching. The distance between the two horn antennas is 115 mm, the height of 3D-printed bracket is 35 mm.

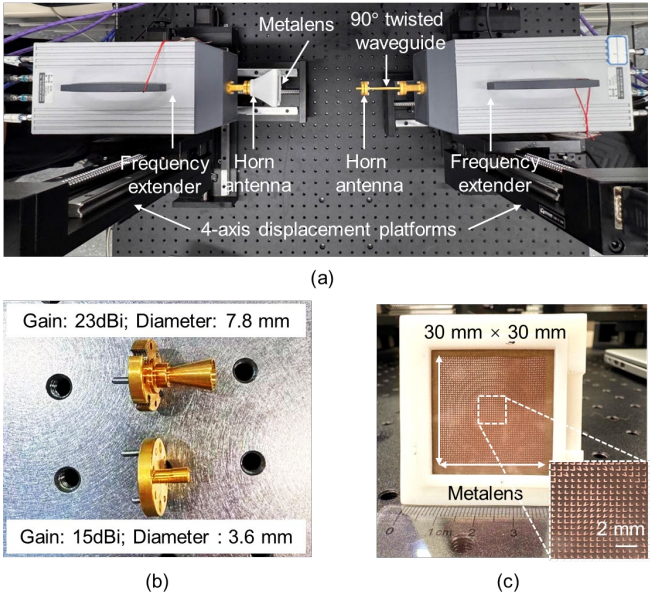

Fig.4 Experimental setups for the measurement of THz metalens antennas. (a) Vector network analyzer using four-axis electric displacement platform. (b) Standard conical horn antennas with gain of 15 dBi and 23 dBi. (c) Photograph of metalens mounted on the 3D-printed bracket.

We employed the 23 dBi gain circular horn antennas and mounted them on the feed ports of expanders. Figure 5(a) depicts the measured received power versus frequency of the 23 dBi antenna and metalens, and the relative gain of the metalens antenna is shown in Fig. 5(b). The involvement of metalens significantly enhances the performance of the standard horn antenna, which already exhibits a peak gain of approximately 33.25 dBi. This peak gain is achieved at a frequency of 0.2469 THz, aligning well with the operational range of both meta-atoms and lens phase distribution. The aperture efficiency ($\eta$) is computed as follows:

$$\eta = \frac{G\lambda^2}{4\pi S} \qquad (5)$$

where $G$ is the measured gain of metalens and described in linear, and $S$ is the size of metalens. In this scenario, the maximum $\eta$ of metalens antenna achieve 27.78%. The 3 dB bandwidth of metalens spans 32 GHz, covering frequencies from 0.228 THz to 0.260 THz. The measured gain at 0.26 THz is still 31.70 dBi, indicating that this metalens antenna may obtain a wider 3 dB bandwidth as the operating frequency increases. Note that conical horn antenna experiences increased harmonic oscillations, which can be attributed to its higher gain and potential device wear over time. To measure the radiation patterns, the motor step of the rotational axis is 0.5° ranging from −30° to +30°, and the transmitter metalens antenna was scanning within this angle range. Figures 5(c1) and 5(c2) illustrate the simulated and experiment radiation patterns in E and H planes at the frequency of 0.24 THz. The experimental data regarding the E-plane and H-plane radiation patterns aligns closely with our simulation results. Additionally, the beam's full width at half-maximum divergence angle is approximately 1.5°. To verify the working bandwidth, measurements of the radiation patterns across a frequency spectrum from 0.20 THz to 0.25 THz were conducted, as shown in Fig. 5(d1) and (d2). The results indicate that the full width at half maximum divergence angles is consistent across the various frequencies. However, we observed that the angles of the side lobes increase as the frequency deviates from 0.24 THz. The increase in side lobe angles is attributed to the defocus effect, which occurs when the frequency strays from the phase designed (0.24 THz) for the lens.

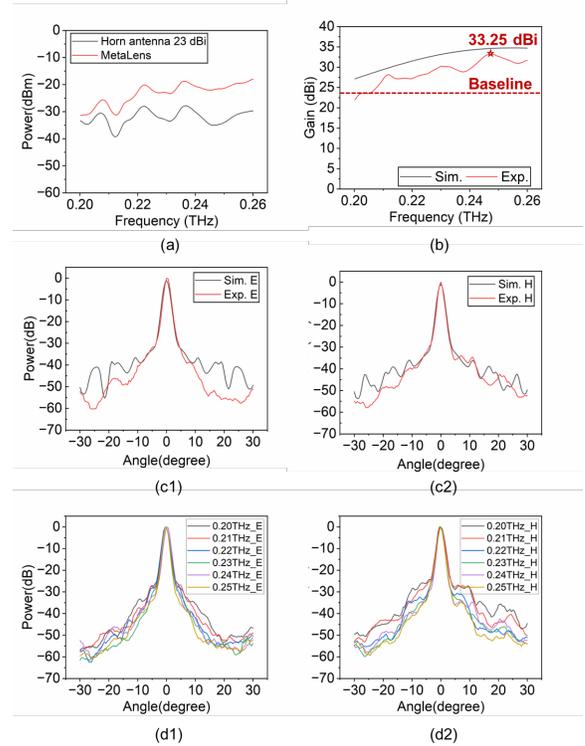

Fig.5 Experimental results of metalens antenna using standard 15 dBi horn antenna. (a) Measured received power versus frequency. (b) Relative gain of metalens antenna versus frequency. (c) Simulated and experimental radiation patterns of (c1) E-plane and (c2) H-plane. Radiation patterns of (d1) E-plane and (d2) H-plane with different frequencies.

*B. Experimental demonstration of THz communication system with metalens antennas.*

As proof of concept, we demonstrate the prototype THz wireless communication system using high gain metalens. The schematic diagram of the wireless communication system and experimental setups are shown in Fig. 6(a) and (b). The key components of the THz radio frequency (RF) chain include sub-harmonic mixers, 8-times frequency multipliers and a THZ low-noise amplifier (LNA). The distance between transmitter and receiver is 260 mm. The bandwidth of the RF chain is 20 GHz from 0.20 to 0.22 THz, which is limited by the sub-harmonic mixer and the low-noise THz amplifier. The clock frequencies of THz transmitter and receiver are the same as 13.475 GHz. The clock frequencies are multiplied by 8-times frequency multipliers to drive the sub-harmonic mixers. The carrier frequency of the THz wireless communication system is 0.2156 THz. On the transmitter side, the sub-harmonic mixer was modulated by using an arbitrary waveform generator. The modulated THz signal from the sub-harmonic mixer was then radiated by the fabricated metalens antenna. On the receiver side, we use the same metalens antenna to receive the modulated THz signal. The low-noise THz LNA with 15 dB gain is put after the sub-harmonic mixer. The demodulated signal was amplified by a 50-GHz-bandwidth 38-dB-gain low-noise amplifier. The digital signal and BER were measured by a real-time oscilloscope.

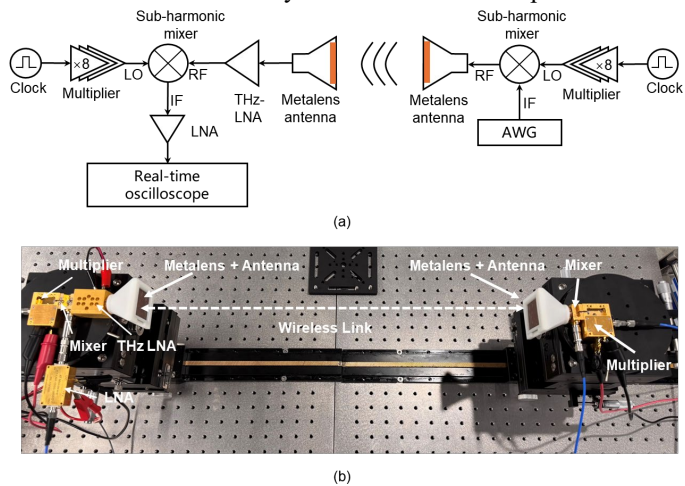

Fig.6 THz wireless communication system. (a) Schematic diagram and (b) experiment setups. LO: Local oscillator; RF: Radio frequency; LNA: Low-noise amplifier; AWG: Arbitrary wavefront generator.

At the receiver end, the electrical signals are captured using a real-time oscilloscope with a sampling rate of 256 GSamples per second. The initial stage of digital signal processing involves applying a frequency shift and a low-pass filter to extract the baseband signal, thereby eliminating out-of-band noise. Subsequently, power normalization and resampling are conducted for signal preprocessing to ensure uniform amplitude levels and to adapt the signal to the processing requirements. A square timing recovery algorithm is utilized for precise time synchronization [34]. To address joint polarization tracking and channel equalization, a radius-directed equalizer is employed [35]. Noise cancellation techniques are implemented to reduce the impact of high-frequency noise. The maximum fast Fourier transform is leveraged to accurately estimate frequency offsets [36]. Furthermore, a blind phase search algorithm is applied to compensate for wireless transmission phase noise.

To verify the performance enhancement provided by the metalens for THz wireless communication system, we implement experiments with and without metalens. The conical horn antennas with 23 dBi gain were mounted on the transmitter and receiver waveguide ports. We evaluated this enhancement from two aspects: modulation baudrate (refer to Fig. 7(a)) and modulation orders (refer to Fig. 7(c)). Firstly, metalens was tested under different baudrate conditions. In this scenario, the baudrate varies from 5 GHz to 20 GHz, using a 16-QAM modulation format, as shown in Fig .4(a). The results demonstrate that there is a significant improvement in the BER performance without bandwidth penalty when the metalens is applied. At a baudrate of 15 GHz, the BER is reduced by three orders of magnitude due to the use of metalens. Notably, the signal bandwidth remains unchanged, as indicated by the amplitude-frequency response of the received signals in Fig. 7(b). That means that higher baudrate can be supported by the metalens. At the target BER of $10^{-4}$, the baudrate is increased from 7.5 GHz to 16.0 GHz with the assistance of metalens. Subsequently, metalens were tested under different modulation orders, as shown in Fig .7(c) and (d). For this scenario, four modulation orders of 2, 4, 5, and 6 with a fixed baudrate of 15 GHz were selected, corresponding to QPSK, 16-QAM, 32-QAM, and 64-QAM format respectively. The results indicate that the metalens improves the BER performance without affecting modulation linearity. As shown in Fig. 7(d), the dispersion of signal clusters is significantly reduced, and the constellation distribution remains unaffected with the assistance of metalens.

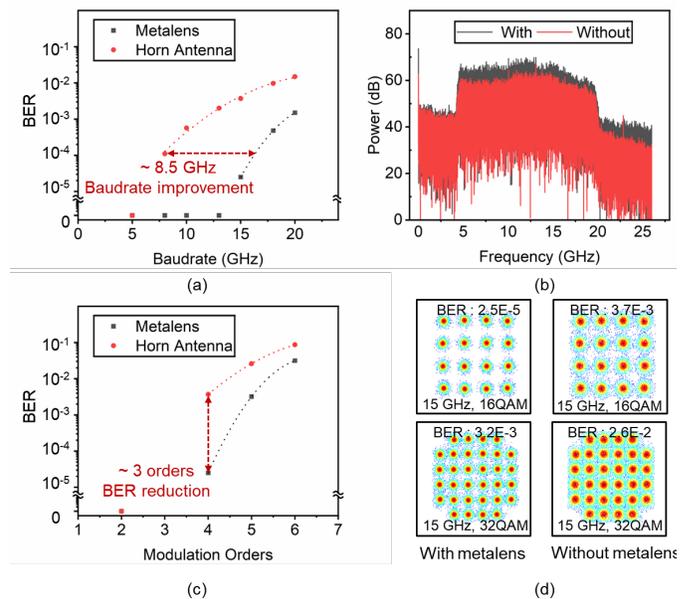

Fig.7 THz wireless communication results for scenarios of modulation baudrate and orders. (a) BER versus baudrate curve. (b) Amplitude-frequency response with 15 GHz baudrate. (c) BER versus modulation curve. (d) Constellation diagrams.

Note that the communication performance significantly deteriorates under high baudrate and high modulation order scenarios. This deterioration occurs because, with constant total input power, increasing the modulation baudrate or the number of formats results in a lower average power per frequency or symbol. As a trade-off, it is crucial to determine the maximum data rate for this system configuration. We have chosen a 32-QAM format with a baudrate of 20 GHz, which approaches the physical limitations of THz transceivers. Under these conditions, the THz wireless communication system achieved a BER of $1.5\times10^{-2}$. Consequently, the system demonstrated a single-channel communication link with a data rate of 100 Gbps. The improvement of communication performance is attributed to the high gain and efficiency of metalens, which can increase the signal-to-noise ratio, preserve signal bandwidth, and maintain modulation linearity. Therefore, metalens emerges as a promising candidate for contributing to the development of high-speed THz communication networks.

*C. Prototype system for real-time uncompressed 4K video transmission.*

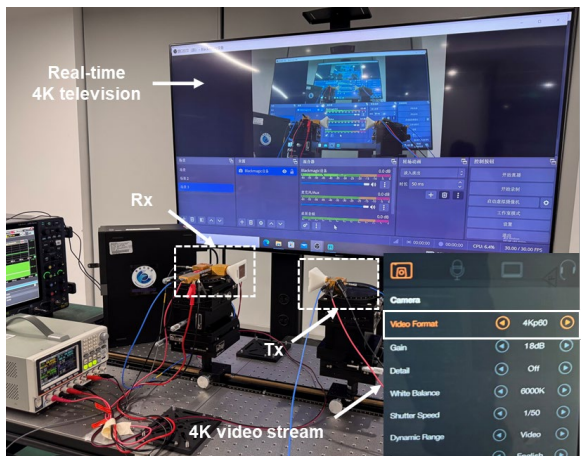

Fig.8 Prototype system for real time transmission of 4K high-definition uncompressed video flow. Refer to Supplementary Video 1 for the demonstration.

To showcase the high-speed transmission capabilities of metalens antenna for 6G communication network, we performed a prototype system for transmitting real-time, uncompressed 4K high-definition video stream, with specifications of 3840×2160 resolution at 60 frames per second and a data rate of 10.24 Gbps. The 4K video stream was captured in real-time using a 4K recorder, and the serial digital interface signals were utilized as baseband signals for the mixer at the transmitter. The signal was then transmitted via a metalens antenna, propagated through wireless channels, and received by another metalens antenna. The received signals were conditioned with an amplifier and converted from SDI to a high-definition multimedia interface signal using a converter card before being displayed on a 4K television. As shown in Fig. 8 and Supplementary Video I, the 4K high-definition video was successfully transmitted and demonstrated in real time.

## IV. DISCUSSION

In the preceding scenario, the high-gain antenna of 23 dBi, characterized by its small waist radius, exhibited a limited aperture efficiency due to its insufficient radiation of electromagnetic waves to metalens. To mitigate this limitation, we use a low-gain horn antenna (15 dBi) and couple with the waveguide port. This integration generates electromagnetic waves with larger waist radius, aligning them more closely with the characteristics of a planar wavefront. Consequently, this adjustment enables the electromagnetic waves to encompass a more extensive region of the meta-atoms within the metalens, thereby significantly enhancing the metalens antenna's gain. Figure 9(a) presents the measured power as a function of frequency for the metalens and the 15 dBi antenna, while the simulated and experiment gains are depicted in Fig. 9(b). The results indicate that this configuration achieves a maximum gain of 36.10 dBi at the frequency of 0.244 THz with maximum $\eta$ of 54.45%. This metalens antenna also achieved an average gain of 34.13 dBi and $\eta$ of 36.71 % with frequency ranging from 0.21 to 0.26 THz. Additionally, lower frequencies outside the designed operational range are attenuated by the metalens, resulting in a 3 dB bandwidth of 33 GHz. The radiation patterns of E- and H-plane are illustrated in Fig. 9(c) and (d). Due to the smaller aperture diameter of the low-gain horn antenna, it exhibits a higher scanning resolution, leading to a reduced half-maximum divergence angle compared to the previous scenario. This maximum gain approaches the physical limitation of metalens with 30 mm × 30 mm regional size. There are two primary methods to increase the gain of a metalens: enlarging the physical size of the metalens and extending the transmission distance. An increased metalens size allows the antenna to capture a more transmitted electromagnetic waves, thereby boosting the received gain of the transceivers. On the other hand, a longer transmission distance can generate a plane wavefront, which illuminates a larger area of the metalens and consequently increases the received gain. It is necessary to clarify that during the THz communication experiment, we utilized a 23 dBi gain antenna instead of the 15 dBi antenna. This configuration was set primarily because the 15 dBi antenna exhibited insufficient signal-to-noise ratio during signal processing, which could not provide comparative analysis against the metalens antenna.

We present a comparative analysis of existing metalenses in the frequency range from 0.15 to 0.30 THz, as detailed in Table II. The evaluation metrics include frequency range, peak gain, peak $\eta$, 3 dB bandwidth, and communication data rates. In terms of modulation strategies, the THz metalenses are categorized into binary (refer to Ref. [24]) and full-phase modulations (refer to [27], [28], [8]). The results reveal that full-phase modulations outperform binary modulations in terms of gain and efficiency, primarily due to the effective suppression of paraxial diffraction in full-phase lens modulations. For experimental demonstrations, metallic planar metasurfaces and 3D-printed structures are commonly employed for fabricating metalenses. The 3D-printed all-dielectric metalens achieves full-phase coverage through the height difference of the cell structure, offering better broadband response due to the absence of metal resonance. However, this approach presents challenges in terms of fabrication complexity and integration. As a trade-off between gain and bandwidth, we propose a planar metalens integrated with standard horn antennas, achieving a gain of 36.1 dBi and a 3dB bandwidth of 33.0 GHz. More importantly, we implemented this integrated device in THz wireless communication, which has not been previously demonstrated within this frequency range. This implementation achieved 100 Gbps wireless communication and 4K uncompressed real-time video transmission, thereby validating its feasibility for 6G communication networks.

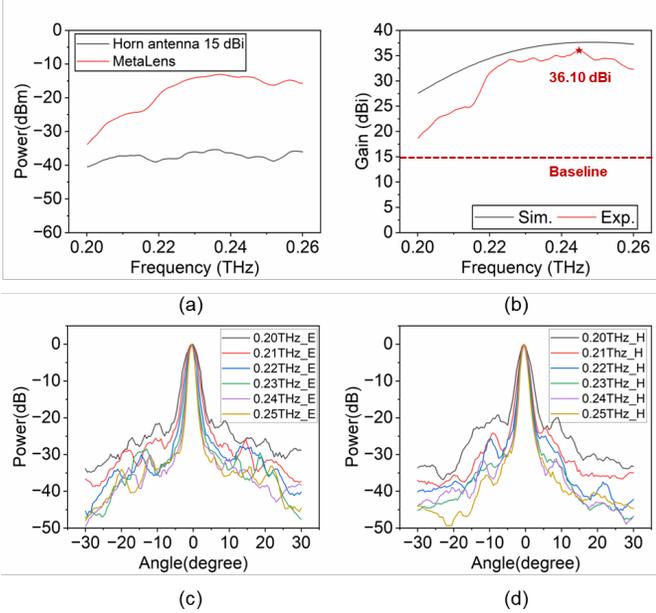

Fig.9 Experimental results of metalens antenna using standard 15 dBi horn antenna. (a) Measured received power versus frequency. (b) Simulated and experimental gain of metalens antenna versus frequency. Simulated and experimental radiation patterns of (c) E-plane and (d) H-plane.

TABLE II
COMPARISON OF THZ METALENS ANTENNAS

| Ref | Freq. (THz) | Size (mm$^2$) | Peak. gain (dBi) | Peak $\eta$ (%) | 3 dB gain bandwidth (GHz) | Data rate (Gbps) |
|---|---|---|---|---|---|---|
| [24] | 0.26−0.28 | 32.0 ×32.0 | 20.8 | 7.0 | < 5.0 | N.A |
| [27] | 0.22−0.28 | 16.74×16.74 | 28.80 | 32.0 | 28 | N.A |
| [28] | 0.25 | 16.8×16.8 | 27.2 | 35.8 | N.A | N.A |
| [8] | 0.26−0.32 | 20.0 ×20.0 | 30.8 | 37.37 | 55 | N.A |
| This paper | 0.20−0.26 | 30.0×30.0 | 36.10 | 56.28 | 33.0 | > 100 |

V. CONCLUSION

We propose a metalens antenna transceiver designed for high gain transmission in THz wireless communication systems. This transceiver is composed of horn antenna with 3D printed bracket and metalens, where the antenna is employed for enhancing the gain of metalens. The metalens employs a layered 'sandwich' architecture that integrates V-shaped copper structure, a dielectric layer, and grating. This configuration allows for resonant enhancement of polarization conversion efficiency and full phase modulation from 0 to 2π.

Our experiments confirm that the metalens achieves a maximum gain of 36.10 dBi and aperture efficiency of 56.28 % with frequency ranging from 0.21 THz to 0.26 THz. Additionally, we have developed a compact THz prototype communication system incorporating these metalens antennas, which has significantly reduced the BER by three orders of magnitude compared to conventional horn antennas. This system also supports a maximum data rate of 100 Gbps (20 GHz baudrate and 32-QAM format), with the BER value below 10$^{-2}$. The proposed metalens provides a compact, high-gain solution for THz wireless communication, potentially accelerating the development of 6G communication network.